\documentclass{llncs}
\usepackage[utf8]{inputenc}
\usepackage{hyperref}
\usepackage{graphicx}

\newcommand{\figref}[1]{figure~\ref{#1}}

\begin{document}

\title{Pervasive Parallelism in Highly-Trustable \\ Interactive Theorem Proving Systems\thanks{The final publication is available at http://link.springer.com.}}
\titlerunning{Paral-ITP}
\author{
Bruno Barras\inst{3} \and
Lourdes del Carmen González Huesca\inst{2} \and
Hugo Herbelin\inst{2} \and
\\ Yann Régis-Gianas\inst{2} \and
Enrico Tassi\inst{3} \and
Makarius Wenzel\inst{1} \and
Burkhart Wolff\inst{1}
}
\authorrunning{Paral-ITP consortium}

\institute{Univ. Paris-Sud, Laboratoire LRI, UMR8623, Orsay, F-91405, France \\
CNRS, Orsay, F-91405, France \and
INRIA, Univ. Paris Diderot, Paris, France \and
INRIA, Laboratoire d’Informatique de l’Ecole Polytechnique
}

\maketitle

\section{Background}

Interactive theorem proving is a technology of fundamental importance for
mathematics and computer-science. It is based on expressive logical foundations
and implemented in a highly trustable way. Applications include huge
mathematical proofs and semi-automated verifications of complex software
systems. Interactive development of larger and larger proofs increases the
demand for computing power, which means explicit parallelism on current
multicore hardware \cite{Sutter:2005}.

The architecture of contemporary interactive provers such as Coq
\cite[\S4]{Wiedijk:2006}, Isabelle \cite[\S6]{Wiedijk:2006} or the HOL family
\cite[\S1]{Wiedijk:2006} goes back to the influential LCF system
\cite{Gordon-Milner-Wadsworth:1979} from 1979, which has pioneered key
principles like correctness by construction for primitive inferences and
definitions, free programmability in userspace via ML, and toplevel command
interaction. Both Coq and Isabelle have elaborated the prover architecture over
the years, driven by the demands of sophisticated proof procedures, derived
specification principles, large libraries of formalized mathematics etc. Despite
this success, the operational model of interactive proof checking was limited by
sequential ML evaluation and the sequential read-eval-print loop, as inherited
from LCF.

\section{Project Aims} 

\begin{figure}[htb]

\centering
\includegraphics[width=0.2\textwidth]{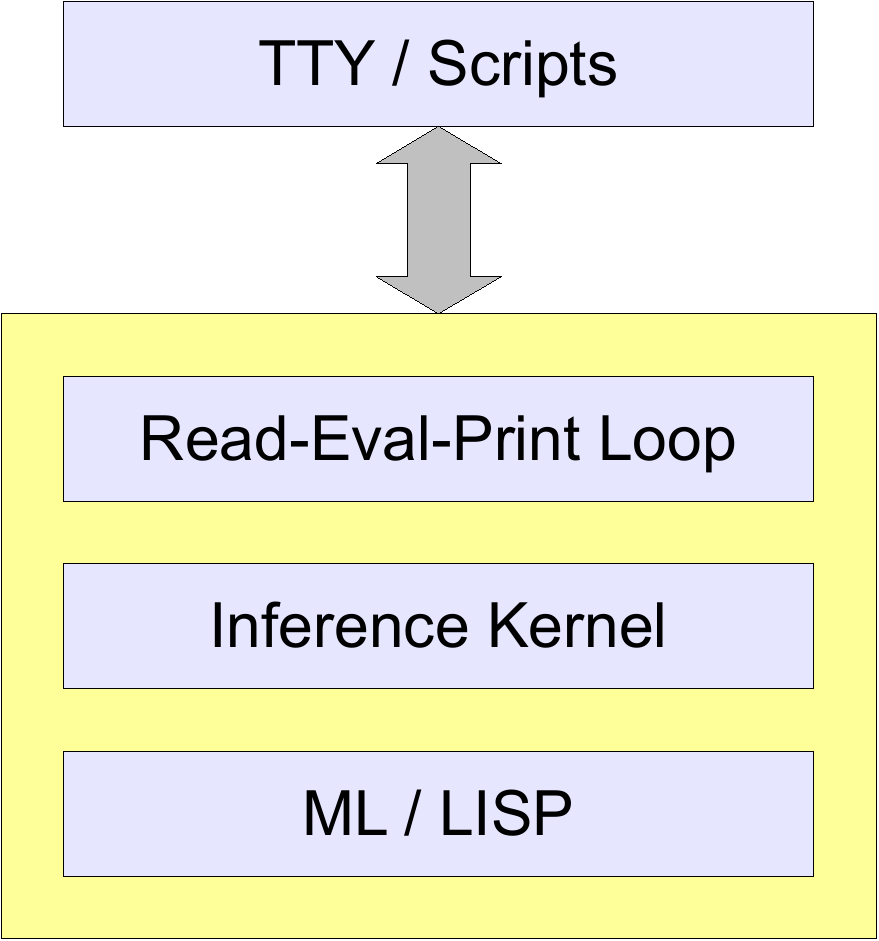} \quad
\includegraphics[width=0.2\textwidth]{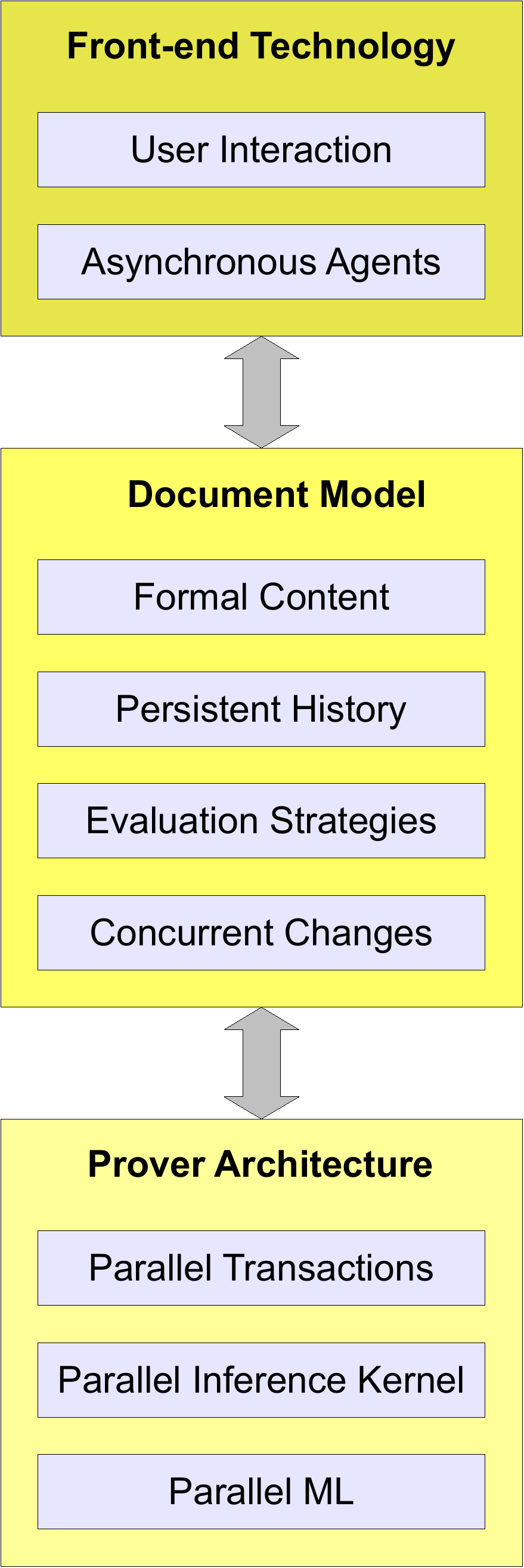}

\caption{Reformed LCF-architecture for parallel proof-document processing}
\label{fig:reformed-lcf}

\end{figure}

The project intends to overcome the sequential model both for Coq and Isabelle,
to make the resources of multi-core hardware available for even larger proof
developments. Beyond traditional processing of proof scripts as sequence of
proof commands, and batch-loading of theory modules, there is a vast space of
possibilities and challenges for pervasive parallelism. Reforming the traditional
LCF architecture affects many layers of each prover system, see
\figref{fig:reformed-lcf}.

Parallelization of the different layers is required on the level of the
execution environments (SML, OCaml), which need to include some form of
multi-threading or multi-processing supported by multi-core architectures.
Isabelle can build on parallel Poly/ML by David Matthews
\cite{Matthews-Wenzel:2010} and earlier efforts to support parallel proof
checking \cite{Wenzel:2009}. For Coq, some alternatives with separate OCaml
processes need to be investigated, because early support for parallel threads in
Caml \cite{Doligez-Leroy:1993} was later discontinued.

Further reforms carry over to the \textbf{inference kernel}, which has to be
extended by means to decompose proof checking tasks into independent parts that
can be evaluated in parallel. The tactic code of proof procedures or derived
specification packages needs to be reconsidered for \emph{explicit} parallelism,
while the inherent structure of the proof command language can be exploited for
\emph{implicit} parallelism. The latter is particularly appealing: the prover
acts like system software and schedules proofs in parallel without user (or
programmer) intervention. Some of these aspects need to be addressed for Coq and
Isabelle in slightly different ways, to accommodate different approaches in
either system tradition.

Our approach is \emph{document-centric}: the user edits a document containing
text, code, definitions, and proofs to be checked incrementally. This means that
checking is split into parallel subtasks reporting their results asynchronously.
The \textbf{document model} and its protocols need to support this natively, as
part of the primary access to the prover process. Finally, a system
\textbf{front-end} is required to make all these features accessible to users,
both novices and experts. Instead of a conventional proof-script editor, the
project aims to provide a full-scale Prover-IDE following the paradigm of
``continuous build --- continuous check''.

%
%
%

These substantial extensions of the operational aspects of interactive theorem
proving shall retain the trustability of LCF-style proving at the very core. The
latter has to be demonstrated by \textbf{formal analysis} of some key aspects of
the prover architecture.

The theoretic foundation of the document model is directed by a
\emph{fine-grained analysis of the impact of changes} made by the user on the
formal text. This analysis not only helps the parallelization of the
verification of the document but also the reuse of already checked parts of the
document that are \emph{almost unimpacted by the user edits}. To give a formal
account on this notion of proof reuse and to implement this mechanism without
compromising the system trustability, we must assign a precise static semantics
to the changes. One foundational part of the project will consist of studying
what kind of logical framework is adapted to the specification and verification
of proof reuses. By the end of the project, we expect to get a language of
semantically-aware and mechanically-verifiable annotations for the document
model.

\section{Current Research and First Results}

Project results are not just paper publications, but actual implementations that
are expected to be integrated into Coq and Isabelle, respectively.  Thus users
of these proof assistants will benefit directly from the project results.

\subsection{A state transaction machine for Coq}

Parallelizing a sequential piece of purely functional code is a relatively
easy task. On the contrary parallelizing an already existing piece of imperative
code is known to be extremely hard.  Unfortunately Coq stores much of its
data in global imperative data structures that can be accessed and modified
by almost any component of the system

For example some tactics, while building the proof, may generate support lemmas on the fly
and add them to the global environment.  The kernel, that will
verify the entire proof once completed, needs to find these lemmas in order
to validate the proof.  Hence distributing the work of building and checking
the proof among different partners is far from being trivial, given that the
lack of proper multithreading in OCaml forces these partners to live in
different address spaces.

In the prototype under implementation \cite{Tassi-Barras:2012} all
side effects have been eliminated
or tracked and made explicit in a state-transaction data structure.
This graph models a collection of states and the transactions needed
to perform in order to obtain a particular state given another one.  
Looking at this graph one can deduce the minimum set of transactions needed
to reach the state the user is interested in, and postpone unnecessary
tasks.  While this is already sufficient to increase the reactivity of the
system, the execution of the tasks is still sequential.  

Running postponed tasks in concurrent processes is under implementation, but we
are confident that the complete tracking of side effects done so far will make
this work possible.

\subsection{Logical framework for semantic-aware annotations}

During the POPLmark challenge, Coq has been recognized as a
metalanguage of choice to formalize the metatheory of formal
languages. Hence, it can semantically represent the very specific
relations between the entities of a proof development. Using Coq as a
logical framework (for itself and for other theorem provers) is
ambituous and requires: (i) to represent partial (meta)programs;
(ii) to design a programming artefact to automatically track dependencies
between computations; (iii) to reflect the metatheory of several logics;
(iv) to implement a generic incremental proof-checker. The
subgoal (i) has been achieved thanks to a new technique of \textit{a
  posteriori simulation of effectful computations} based on an
extension of monads to \textit{simulable}
monads~\cite{Claret:2013:ITP}. The goal (ii) is investigated through a
generalization of adaptative functional programming~\cite{Acar:2006}.

\subsection{Parallel Isabelle and Prover IDE} The first stage of multithreaded
Isabelle, based on parallel Poly/ML by David Matthews, already happened during
2006--2009 and was reported in \cite{Wenzel:2009,Wenzel:2012:PIDE}. In the
project so far, the main focus has been improved scalalibity and more uniformity
of parallel batch-mode wrt.\ asynchronous interaction. Cumulative refinements
have lead to saturation of 8 CPU cores (and a bit more): see
\cite{Wenzel:2013:ITP} for an overview of the many aspects of the prover
architecture that need to be reconsidered here.

The Isabelle2011-1 release at the start of the project included the first
officially ``stable'' release of the Isabelle/jEdit Prover IDE
\cite{Wenzel:2012:PIDE}, whose degree of parallelism was significantly improved
in the two subsequent releases Isabelle2012 (May 2012) and Isabelle2013
(February 2013). The general impact of parallelism on interaction is further
discussed in \cite{Wenzel:2012:UITP}.

Ongoing work investigates further sub-structural parallelism of proof
elements, and improved real-time reactivity of the
implementation. Here the prover architecture and the IDE front-end are
refined hand-in-hand, as the key components that work with the common
document model.  The combination of parallel evaluation by the prover
with asynchronous and erratic interactions by the user is particularly
challenging.  We also need to re-integrate tools like
Isabelle/Sledgehammer into the document model as \emph{asynchronous
  agents} that do not block editing and propose results from automated
reasoning systems spontaneously.

\subsection{Prover IDE for Coq} Once that the Coq prover architecture
has become sufficiently powerful during the course of the project, we
shall investigate how the Isabelle/PIDE front-end and Coq as an
alternative back-end can be integrated to make a practically usable
system. Some experiments to bridge OCaml and Scala in the same spirit
as for Isabelle have been conducted successfully
\cite{Wenzel:2013:CoqPIDE}.  An alternative (parallel) path of
development is to re-use emerging Prover IDE support in Coq to improve
its existing CoqIde front-end, to become more stateless and timeless
and overcome the inherently sequential TTY loop at last.

\section{Project Partners} 

The project involves three sites in the greater Paris area:

\begin{itemize}

\item The \emph{LRI ForTesSE} team at UPSud (coordinator: \textbf{B. Wolff}),
including members from the \emph{Cedric} team (CNAM),

\item the \emph{INRIA Pi.r2} team at PPS / UParis-Diderot (site leader:
\textbf{H. Herbelin}), including members from the \emph{INRIA Gallium} team, and

\item the \emph{INRIA Marelle-TypiCal} team at LIX / Ecole Polytechnique (site
leader: \textbf{B. Barras})

\end{itemize}

Research is supported by under grant \emph{Paral-ITP
  (ANR-11-INSE-001)} with formal start in November 2011 and duration
of 40 months total.  Further information is available from the project
website \url{http://paral-itp.lri.fr/}.

\bibliographystyle{splncs_srt.bst}
\bibliography{root}

\end{document}